# Exploring University Students' Engagement with Online Video Lectures in a Blended Introductory Mechanics Course

Shih-Yin Lin [1], John M. Aiken[2], Daniel T. Seaton[3],
Scott S. Douglas[2], Edwin F. Greco[2], Brian D. Thoms[4], Michael F. Schatz[2]

[1] Department of Physics, National Changhua University of Education, Changhua, Taiwan 500
[2] School of Physics, The Georgia Institute of Technology, 830 State Street, Atlanta, GA USA 30332
[3] VPAL Research, Harvard University, Cambridge, MA 02139 Harvard University http://vpal.harvard.edu/research
[4] Department of Physics and Astronomy, Georgia State University, Atlanta, GA USA 30303

The advent of MOOCs has stimulated interest in using online videos to deliver content in university courses. We examined student engagement with 78 online videos that we created and were incorporated into a one-semester blended introductory mechanics course at the Georgia Institute of Technology. We found that students were more engaged with videos that supported laboratory activities than with videos that presented lecture content. In particular, the percentage of students accessing laboratory videos was consistently greater than 80% throughout the semester while the percentage of students accessing lecture videos dropped to less than 40% by the end of the term. Moreover, students were more likely to access the entirety of a laboratory video than a lecture video. Our results suggest that students may access videos based on perceived value: students appear to consider the laboratory videos as essential for successfully completing the laboratories while students appear to consider the lecture videos as something more akin to supplementary material. We found there was little correlation between student engagement with the videos and the performance in the course. In addition, an examination of the in-video content suggests that students focus more on concrete information that is explicitly required for assignment completion (e.g., actions required to complete laboratory work, or formulas/mathematical expressions needed to solve particular problems) and less on content that is considered more conceptual in nature. The results of the study suggest ways in which instructors may revise courses to better support student learning.

## I. INTRODUCTION

With the advent of new online educational technologies and the popularity of MOOCs (Massively Open Online Courses), there is increasing interest in leveraging web-based resources in university courses. Using video lectures, online forums, social networking applications, etc., instructors are finding new ways to engage students both in class and outside of class [1-5]. The MOOC movement has also invigorated interest in data-driven education [6-10], particularly due to the sheer size and scope of data being collected by platforms, which are able to provide second-by-second records of student engagement with resources in a course. Such data promise to provide educational researchers powerful and unprecedented insight into student learning behaviors [6-8]. For example, by analyzing how students engage online content, educators can begin to paint a picture of what resources students attend to and how they attend to these resources [6]. The effects of these online resources on student learning outcomes can also be explored.

In 2013, video lectures that we originally created for a MOOC were implemented in a blended introductory mechanics course at the Georgia Institute of Technology. In order to save class time for group collaborative work, the traditional in-class lectures were replaced by online videos that students were instructed to access outside of class at their convenience. As part of an effort to investigate student participation and learning in this new type of course, we explored student engagement with online videos.

In our course, the videos were hosted on the Coursera MOOC platform, which records not only whether a student accesses a video, but also every interaction a student makes with their video player in a tabulated timestamped output (e.g., pauses, plays, and seeks, which indicate when a student is skipping parts of the video to find a segment of interest). These records allow educators to investigate how students engage with videos in numerous ways. For the goal of our paper, we focus on the following five distinct types of analysis, with relevant research questions presented below:

(1) Video-Accessing Sessions (i.e., whether students clicked on the links to access videos)
   Fall 2013 was the first time in-class introductory mechanics lectures were replaced by online lecture videos at the Georgia Institute of Technology. Therefore, understanding student use of this new online resource is of interest to instructors and researchers. Research questions in this realm include: *To what extent did students access the video lectures? Did they access the videos multiple times or did they access each video only once? What videos (if any) did students access more?*

In this paper, we use the term *starting a video-accessing session* to refer to the action of a student *clicking on the link* for a video to access the content through online streaming or to download it. Note that if the same student clicks on the link for the same video multiple times, multiple *video-accessing sessions* are made. The *number of unique videos accessed* by a given student, on the other hand, refers to how many *different* videos he/she had ever accessed. It does not take into account how many times any given video is accessed by the same student.

(2) Timing of video access

Research questions in this realm include: *How do the students' video-accessing behaviors relate to the course schedule (e.g., the dates when videos were assigned, the due dates for homework or laboratory work.)* A description of the course structure will be presented in Section III.

(3) Fraction of in-video content accessed

While the first two types of analysis focus on whether and when students clicked on the video links, the third type of analysis explores how students interacted with the videos after the video started playing. In particular, we focus on the following question:

*When students accessed videos, did they access the video completely, or did they access only parts (but not all) of the videos? More specifically, for each video, what fraction of students accessed "almost" all of the video?*

(4) Relation between video accessing and student performance

With an understanding of how students in the course accessed videos, we then investigate *if students' performance in the course is correlated to their video accessing behaviors (such as the number of videos accessed, the time spent watching video, or the interaction frequency with video player).*

(5) Detailed student interaction with videos

In order to help understand the relation between video accessing and student performance further, an exploration into *students' detailed interaction* with two selective videos (e.g., *what types of content did students seem to engage more with?*) is discussed.

Before we proceed, we would like to remind the readers that in this paper, terms like "accessing", "engagement", etc., represent student interaction with the videos *as recorded by the video player*. Due to the limitation of data available – we only have access to the timestamped clickstream data recorded by the video player but cannot monitor how attentive students were when the videos were playing –, the *extent* to which students were engaged when playing a video is beyond the scope of this study. In the remainder of the paper, we first discuss existing studies relevant to student engagement with videos, and then provide a detailed description of the course structure and the video lectures. We then describe our research methodology, report findings from the study, and conclude with possible future work that may have the potential to help improve student learning in the course.

## II. BACKGROUND

With the increasing use of videos in educational settings, student learning via lecture videos (whether presented as supplemental material or as the primary means of delivering course content) has gained interest to the educational community. One thread of research commonly explored by educators is the extent to which students make use of these resources [7] and how video accesses correlate to student performance in the course [11,12]. This research has produced mixed results. For example, a study in an European Law course [11] in which video recordings of in-class lectures were provided as supplemental materials shows that when controlled for important factors such as GPA and time spent studying, the number of post-class video-accessing is positively correlated to students' course grade. On the other hand, another study [12] in two calculus courses shows that the tendency to both attend in-class lectures and access online recording of previous lectures is negatively correlated to students' final grade. Similar mixed results are not only reported at the video-accessing session level, but also at the detailed interaction level. For example, researchers from the University of Toronto at Scarborough have found that while the usage of pauses and seeks in videos is related to higher exam scores in an introductory psychology course, usage of pauses is negatively correlated to final grade in a calculus course [12]. These studies suggest that many factors can influence the relationship between video accessing and student performance. For example, while repeated access allowed by videos may help improve students' understanding of the materials, it is also possible that students who need to access the videos most are those who have greater difficulty in the course. Moreover, factors such as the types of content covered in the videos (e.g. communication of concepts vs. procedural skills to solve problems), and the learning strategies used by students (e.g. maximizing understanding vs. avoiding failure with minimum time and effort) can all affect the relationship between video accessing and student performance [12]. To our knowledge, few studies have reported on how video accessing relates to student learning in an introductory physics course [13], especially when the videos were provided as major resources that introduce students to important concepts and skills in the course. The current study attempts to explore this issue in the context of our on-campus, blended course setting.

In addition to the connection between video accessing and learning outcomes, student engagement of lecture videos [7,8] itself is also of interest because understanding how students use lecture videos can help instructors design better lecture videos. If peaks are observed in students' accessing trajectory of in-video content, places where these peaks occur may indicate what students consider to be important points of interest in the given video. On the other hand, if students commonly disengage with specific points of a video, changes may be made to the video to help students benefit from it more. A study of user interaction with hundreds of videos from four edX MOOCs [8] identifies 5 patterns for peaks in students' in-video content accessing, which are most often due to visual transitions such as a video beginning new material, students returning to watch missed content, students pausing and leaving the video to complete a tutorial step, or students replaying a small segment of the video surrounded by visual transitions both before and after. The same study also finds that students accessing videos often do not complete videos. A predicted dropout rate of 53% or more is obtained for videos exceeding 5 minutes long using a linear regression model [8]. Similar work on student engagement with videos has led to a series of suggestions on future video production [8,10]. For example, videos should be short, and should avoid abrupt transitions. Moreover, interactive links or timelines help students find common points of interests in their re-access of videos. We note that in the prior study discussed here, dropout is defined as the last point of access in a video accessing session, regardless of what students might have accessed or skipped through earlier in the given video. Since students can access the video non-linearly, in our current work, the relation between video length and student-video interaction is explored from a slightly different angle in terms of what fraction of in-video content is accessed by the students. In addition, we explore this issue further by comparing the results between videos of different content types.

## III. COURSE STRUCTURE AND THE VIDEO LECTURES

### 1. Out-of-class Activity

The "blended" introductory mechanics course explored in this study was taught with the "Matter and Interactions" [14] curriculum. This curriculum, which places an introductory physics course in a modern context and emphasizes important scientific practices like modeling and computation, has been offered at the Georgia Institute of Technology in a large-lecture course for many years. In order to engage students in more group collaborative work during the in-class time, starting in Fall 2013, a "blended" version of this course was offered in which the traditional lectures were substituted by online videos. In particular, a number of lecture videos we created were assigned each week. Students were instructed to watch these online videos before coming to class for group problem-solving work. They were also instructed to perform laboratory activities individually at home by observing the motions of objects in their own surroundings, analyzing these motions through video analysis [15-17], and modeling the motions via Python programming language. A total of five laboratory activities were implemented in this course, and the final project for each laboratory activity is to produce a short video report detailing their work and their results. Each laboratory activity had a 2-week cycle, with students performing the laboratory activity and creating a lab report in the first week, and peer-evaluating each others' reports in the $2^{nd}$ week. Other out-of-class activities students participated in involved homework assignments, textbook readings, and online forum discussion.

### 2. In-class sections

In total, 161 students enrolled in the blended course, and they were split into sections of approximately 25 students each. Every week, students met in-class with the instructor and the TAs in their own section for 3 hours. About 2 hours were spent on group problem solving in which students worked on tutorial-style problem-solving worksheets with 2 or 3 peers at their table. During this period, the instructor and TAs circulated the class to interact with the students and to assist students with their work when needed. About one hour of class time was spent on lab presentations, in which students practiced presenting their laboratory work to their peers either in the form of a draft report (during the first week of the lab cycle), or a final report (during the $2^{nd}$ week of the lab cycle). The lab presentation section was led by a teaching assistant or the course instructor who would provide feedback and guide discussions among the presenter's peers about how to best meet the goals of their presentations. After students participated in the in-class sections, which were held on Monday to Thursday, a weekly quiz was held on Friday to allow students to check for their understanding of the materials. These quizzes were conducted in a proctored setting and administered on computers. There was one written midterm exam and one written final exam during this 17-week-long course.

### 3. Video Lectures

As described earlier, students were required to participate in lectures and laboratory work outside of classroom at their convenience. A total of 78 video lectures were assigned throughout the whole semester to assist students with these activities, with the number of videos assigned each week shown in Figure 1. These 78 videos can be grouped into two categories: 64 of them were "lecture-oriented" videos that introduced students to specific physics concepts and/or problem solving skills. These videos covered contents that an instructor would typically discussed in lectures, most of which were whiteboard animated with the intent to attract and to hold the interest of students. The other 14 videos were "laboratory videos", which were tied to skills and concepts necessary for successfully completing the at-home laboratory activities. In particular, 8 of them were "lab-specific videos" that provided specific information relevant for completing a particular lab. Six of them were "supplemental laboratory videos" which introduces general concepts, skills or techniques that were generally useful for all laboratory activities. Table I provides a few example video topics in each group. All 78 videos were short, typically 5-20 minutes long. Students can access these videos by streaming them online as well as downloading them. These videos make up the data explored in the rest of this paper.

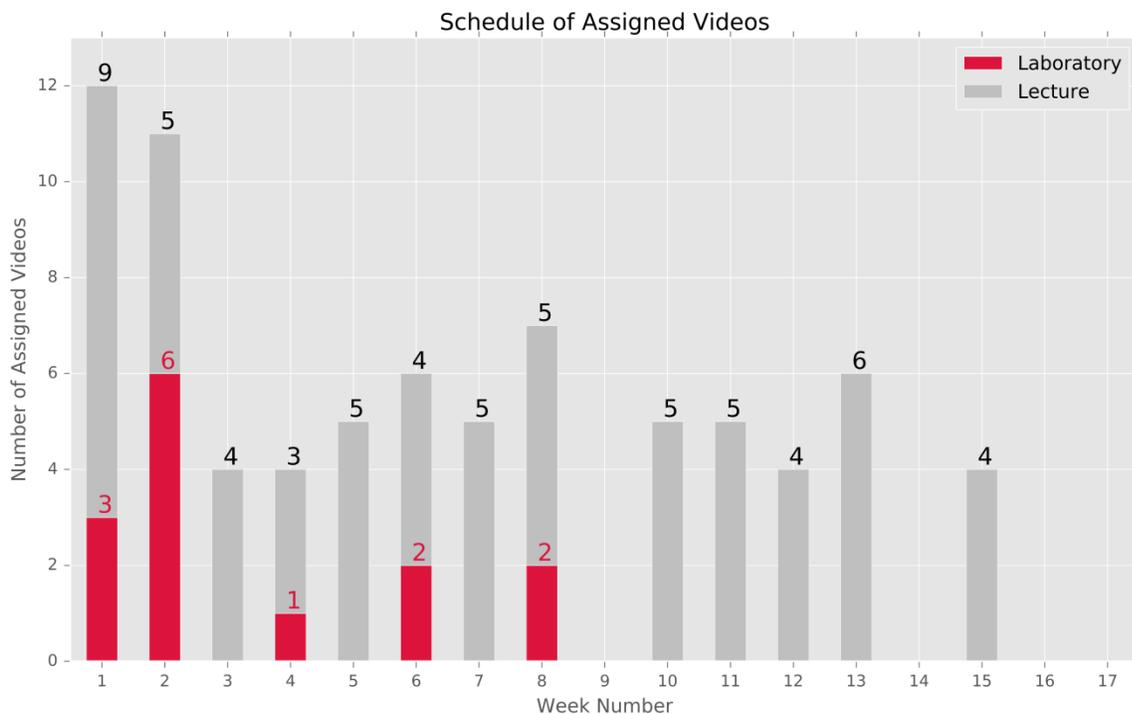

Figure 1: (Color online) Number of videos assigned each week. Laboratory videos specific for the 1st, 2nd, 3rd, and 4th lab were assigned in weeks 1&2, week 4, week 6, and week 8, respectively. The 5th lab was a "choose your own adventure" lab and students were expected to take advantage of what they learned in the course to explore any kind of motion that were of interest to them. Therefore, no laboratory videos specific for this lab was provided. Week 1 and 2 included supplemental lab videos concerning specific tools used in all lab activities instead of specific lab activities.

Table I. Examples of videos in each category

| Category | Examples |
|---|---|
| "Lecture-oriented" Videos (N=64) | • Vectors in 1-D<br>• Newton's $2^{nd}$ Law<br>• Spring Potential Energy |
| "Laboratory Videos" (N=14) | *Lab- Specific*:<br>• Video Analysis of Constant Velocity Motion: How to use Tracker<br>• Creating a Computer Model of Constant Velocity Motion<br>• Black Hole Lab Introduction<br>*Supplemental*:<br>• Installing VPython<br>• Using a Spreadsheet<br>• Recording Observations on Video<br>• Creating a Good Video Lab Report |

## IV. METHODOLOGY

To provide a sense of how analysis related to student-video interaction is conducted, we first present a single video-streaming session by a single student. Then, analysis of student aggregate behaviors, which is the focus of this paper, will be discussed.

### 1. Example from a single streaming session by a single student

Figure 2 shows an example of a student's video streaming behavior. In particular, this was the fifth time student #3 accessed video #15 through online streaming; this was also the first time this student accessed the given video to its conclusion. Three types of events recorded of student interaction are shown in this figure: plays, pauses, and seeks. Plays (represented by gray triangles in Figure 2) and pauses (represented by red squares) can be manually generated by the student or auto-generated by the video player. For example, there will always be a play at the beginning of each video and a pause at the end. Seeks are recorded when a student clicks on the scrubber below the video to skip to a different point in the video. If the video player reaches an interaction point (represented by vertical dashed lines in Figure 2), a pause is auto-generated, and a new window pops up asking students to complete a task (e.g., downloading a supplemental file or answering a question.) When students exit the interaction point to access the rest of the video, a play will be recorded. On Coursera, several playback rate options (from 0.5x to 2x, in increments of 0.25) are available so that students can watch the video at their preferred rate. The dashed diagonal line in Figure 2 represents where "video time" equals "real time." Thus, if the slope between a play event and the subsequent pause event is larger (smaller) than the slope of the dashed diagonal line, the playback rate students used is larger (smaller) than normal playback of 1.0x. This video is 1052 seconds long (~17.5 minutes).

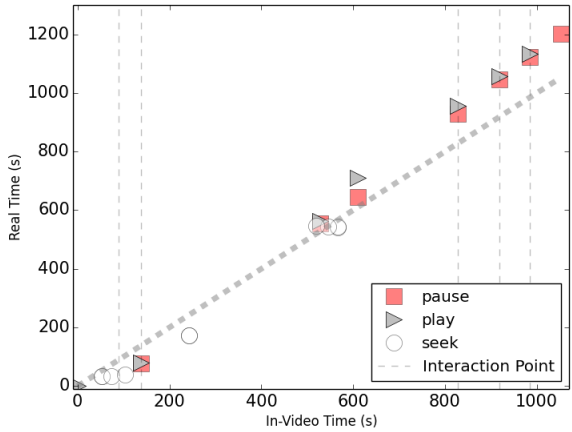

Figure 2. The "accessing trajectory" of a single student. The dashed diagonal line represents where video time and real time are identical.

For each student-video interaction, plots like Figure 2 provide an "accessing trajectory", that is, a snapshot of how each student interacted with the video. Beginning from the origin of Figure 2, this student played the video for approximately 32 seconds in real-time and then began to seek through the video. Because the video did not pause at the first interaction point (represented by the first vertical dashed line), this provides evidence that the student skipped past this interaction point. The video auto-paused at the second interaction point, and playback was resumed ~2.7 seconds later. The student then played the video for ~91 seconds, skipped slightly ahead to time 242 s in the video, and then continued playing ~370 seconds before skipping to pause at time 528 s in the video. The student then played the video for another ~82 seconds until pausing for ~64 seconds at time 610 s in the video. Subsequent pause-play pairs correspond to the remaining three interaction points in the video. The majority of play-pause pairs in Figure 2 have a slope equal to that of the dashed diagonal line, suggesting that this student watched the video with normal playback rate (1.0x) most of the time.

## 2. Analysis of students' aggregate video-accessing behaviors

For each video-streaming session, an accessing trajectory similar to that discussed above can be retrieved from the data. With these accessing trajectories available, we can explore how students as a group accessed videos. To facilitate discussion of how our analysis is performed, we first introduce a matrix $A_{ijkt}$, with the indices correspond to the following:

- $i$, student ID (it takes values from 1 to $N_s$, where $N_s$ represents the total number of students. In our study, $N_s =161$),
- $j$, video number in the assigned order (it takes values from 1 to $N_v$, where $N_v$ represents the total number of video lectures in the course. $N_v =78$),
- $k$, index of streaming sessions, i.e., the $k$th time a given student $i$ clicked on the link for video $j$ to access its content through online streaming ($k$ takes values from 1 to $N_{a,max}$, where $N_{a,max}$ represents the maximum number of times a student in our course has clicked on the link for the same video to stream it repeatedly. In our study, $N_{a,max} = 25$ ), and
- $t$, in-video time in second ($t$ takes values from 0 to $t_j$, where $t_j$ represents the length of the given video $j$ in seconds. While Coursera records data in the millisecond accuracy, for the purpose of this paper, we use 1 second as the sampling rate interval for simplicity);

For each video-streaming session (i.e., a given set of ($i, j, k$)), we construct a matrix representing how many times student $i$ has accessed the content at in-video time $t$ of video $j$ during the $k$th time she streamed this video online; a value of 1 is added to element $A_{ijkt}$ every time the designated point of the video was accessed; otherwise, a value of 0 is assigned. Figure 3 presents a slice of the matrix $A_{ijkt}$ for the example shown in Figure 2 (with, i=3, j=15, k=5 since this was the 5th time student 3 streamed video 15). A detailed discussion of how the matrix $A_{ijkt}$ is calculated from the play, pause, and seek data recorded can be found in the Appendix. We note that if a given student $i$ has never accessed a given video $j$ through online video streaming, $\sum_{kt} A_{ijkt} = 0$. On the other hand, $\sum_{kt} A_{ijkt} \neq 0$ means that the student has clicked on the link for video $j$ at least once to access its content online. We also note that in addition to streaming videos online, students could also download videos to access them offline. A total of 17092 cases of video streaming and 145 cases of video downloading were recorded in this course. Due to the types of data available, array $A_{ijkt}$ are constructed only for the former cases of online video streaming. For the latter cases of video downloading, another matrix $D_{ij}$ was constructed to represent how many times student $i$ has downloaded video $j$. $D_{ij}$ can be greater than one because some students were found to download the same video multiple times.

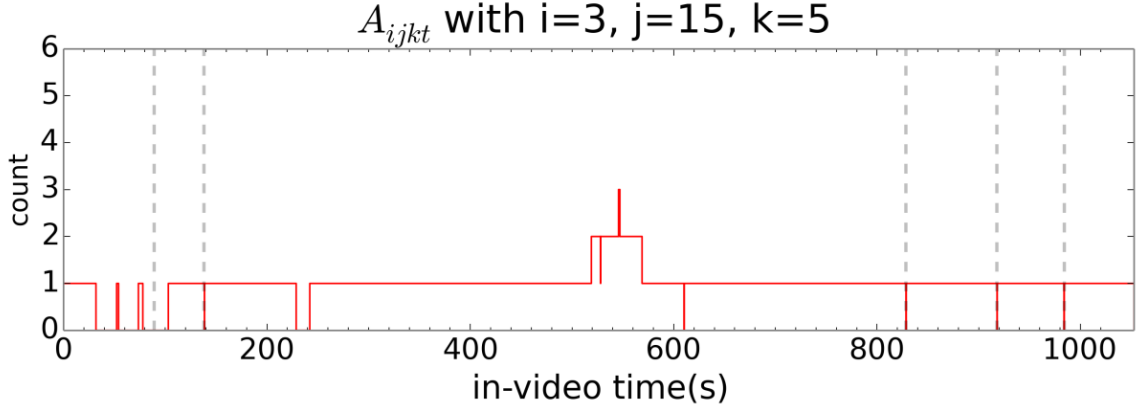

Figure 3. Time series indicating the number of times student 3 accessed a given second in video 15 during that student's 5[th] accessing session. In this session, student 3 accesses most of the video once while accessing 519s ~567s multiple times.

Using $A_{ijkt}$ and $D_{ij}$, we can define a few more quantities to represent students' video accessing behaviors from different aspects:

(1) $TS_{ij}$: Total number of sessions (i.e., how many times student $i$ has clicked on the link for video $j$ to either access its content online or to download it)

$$TS_{ij} = TS_{ij\ from\ streaming} + TS_{ij\ from\ downloading}$$
$$= \begin{cases} \max(k) & | \sum_t A_{ijkt} > 0 \\ 0 & if \sum_t A_{ijkt} = 0 \text{ for any given } k \end{cases} + D_{ij}$$

(2) $US_{ij}$: Unique session (i.e., whether student $i$ has ever clicked on the link for video $j$ to access its content through online streaming or downloading)

$$US_{ij} = \begin{cases} 1 & if\ TS_{ij} \neq 0 \\ 0 & if\ TS_{ij} = 0 \end{cases}$$

(3) $V_i$: Fraction of unique videos accessed by a student (i.e., what proportion of all videos in the course ($N_v = 78$) student $i$ has ever accessed)

$$V_i = \frac{\sum_{j=1}^{N_v} US_{ij}}{N_v}$$

When further separating the videos into lecture-oriented videos (64) and laboratory videos (14), similar fractions can be defined for videos with different content type:

$$V_{i,lec} = \frac{\sum_j US_{ij}\ \text{for}\ j \in \text{Lecture} - \text{oriented videos}}{64}$$
$$V_{i,lab} = \frac{\sum_j US_{ij}\ \text{for}\ j \in \text{Laboratory videos}}{14}$$

(4) C(V): Complimentary cumulative distribution function of the amount of unique videos accessed by students (i.e., what fraction of students have accessed more than a certain proportion of videos in the course. Here, V represents the fraction of videos in a category). In particular,

$$C(V) = \frac{Card(\{V_i | V_i > V\})}{N_s}, 0 \leq V \leq 1$$

where $Card()$ represents the cardinality, i.e., number of elements, of a given set.

When further separating the videos into lecture-oriented videos and laboratory videos, similar functions can be defined for videos with different content type:

$$C(V)_{lec} = \frac{Card(\{V_{i,lec} | V_{i,lec} > V\})}{N_s}, 0 \leq V \leq 1$$

$$C(V)_{lab} = \frac{Card(\{V_{i,lab} | V_{i,lab} > V\})}{N_s}, 0 \leq V \leq 1$$

(5) $NS_j$: Number of students who has ever accessed a given video $j$ through online streaming or downloading

$$NS_j = \sum_{i=1}^{N_s} US_{ij}$$

(6) $FS_j$: Fraction of students accessing a video (i.e., out of all the students in the course, what proportion of students have ever accessed the given video $j$)

$$FS_j = \frac{NS_j}{N_s}$$

(7) $FMS_j$: Fraction of students with multiple accessing sessions for a given video (i.e., out of all the students who accessed a given video $j$, what proportion of them have accessed it more than once)

$$FMS_j = \frac{Card(\{TS_{ij}|TS_{ij} \geq 2, i = 1 \sim N_s\})}{Card(\{TS_{ij}|TS_{ij} \geq 1, i = 1 \sim N_s\})}$$

(8) $TA_{ijt}$: Total number of accesses for specific in-video content by a single student (i.e., for a selected video $j$, how many times has student $i$ accessed the content at $t$ second of the given video. Due to the availability of data, this variable as well as the next three variables that follow are constructed only for cases of online video streaming. The few cases of video downloading [18] were excluded in such type of analysis.)

$$TA_{ijt} = \sum_{k=1}^{\max(k)} A_{ijkt}$$

(9) $UA_{ijt}$: Unique access for specific in-video content by a single student (i.e., for a selected video $j$, whether student $i$ has ever accessed the content at $t$ second of the given video)

$$UA_{ijt} = \begin{cases} 1 & \text{if } TA_{ijt} \neq 0 \\ 0 & \text{if } TA_{ijt} = 0 \end{cases}$$

(10) $FC_{ij}$: Fraction of in-video content accessed (i.e., total fraction of video $j$ that student $i$ has ever accessed)

$$FC_{ij} = \frac{\sum_{t=1}^{t_j} UA_{ijt}}{t_j}$$

(11) $FSE_j$: Fraction of students accessing "almost" the entirety of video $j$ (i.e. out of all the students who accessed a selected video $j$, what proportion of them have accessed "at least 99%" of the content in the given video. Here, the criteria of accessing the video "almost" entirely instead of 100% completely is applied because in some situations, students may think they have received all the information presented in a video even though strictly speaking not every single second of the given video has been accessed. For example, students may skip the opening music at the beginning of every video because they find it irrelevant to physics. In addition, students may require less time than provided to digest the information presented and therefore stop a video one or two seconds earlier than its official ending when only static footage but not audio was presented at the very end.) In this paper, we set the criteria of accessing "almost" the entirety of a video as accessing "at least 99%" of the given video. $FSE_j$ is defined as

$$FSE_j = \frac{Card(\{FC_{ij}|FC_{ij} \geq 0.99, i = 1 \sim N_s\})}{NS_{j,streaming}}$$

where $NS_{j,streaming}$ represents the total number of students who has ever accessed video $j$ through online streaming and is defined as

$$NS_{j,streaming} = \sum_{i=1}^{N_s} US_{ij,streaming} \quad \text{where}$$
$$US_{ij,streaming} = \begin{cases} 1 & \text{if } TS_{ij,from\ streaming} \neq 0 \\ 0 & \text{if } TS_{ij,from\ streaming} = 0 \end{cases}$$

We note that while the first seven quantities presented here focus on whether students have ever clicked on the link for a given video to download it or to access its content online (regardless of how much in-video content was accessed), the last four quantities focus on the exact in-video content that was accessed by students. A short summary of the quantities discussed in this section can be found in Table II.

## V. RESULTS

### 1. Unique Sessions

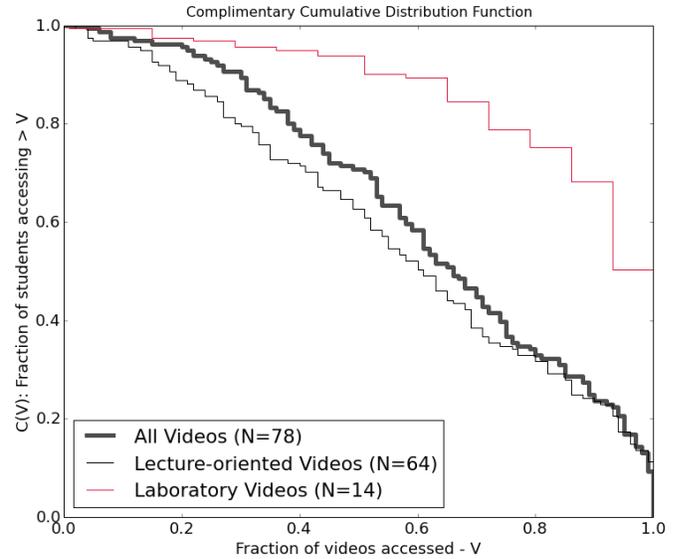

Figure 4. Complimentary Cumulative Distribution Function C(V) showing the fraction of students accessing more than a certain proportion of videos in each group Students are much more likely to access most or all of the laboratory videos in comparison to the lecture-oriented videos.

The first major goal of this paper addresses the issue of to what extent students made use of the video lectures online. Figure 4 presents the complementary cumulative distribution functions of the unique videos accessed by students for each type of videos (i.e., C(V), C(V)$_{lec}$, C(V)$_{lab}$ defined earlier in the methodology section). Overall, less than 20% of students accessed all 78 videos, and half of the students skipped more than 35% of the videos, suggesting that students did not think it is necessary to view all of the videos exhaustively. A comparison between the fraction of students accessing lecture-oriented videos and the fraction of students accessing laboratory videos shows that lab videos were accessed more than the lecture-oriented videos. While more than half of the students accessed all the lab videos, less than 20% of the students accessed all the lecture-oriented videos.

Table II. Short summary of the quantities used for describing students' video-accessing behaviors. In this table, quantities presented in the white background are obtained by examining whether or how many times students have started a video-accessing session (i.e., clicking on the link for a video) regardless of the amount of in-video content accessed within each session. On the other hand, quantities presented in the shaded background are obtained by focusing on the exact in-video content that was accessed by students. In addition, the ► symbol indicates quantities that are constructed for cases of video accessing through online streaming, while the ↓ symbol indicates quantities that are constructed for cases of video accessing through downloading.

| Type | Quantity | Source of Data | Meaning |
|---|---|---|---|
| Fixed Quantities | $N_s$ | | Total number of students in our study. $N_s = 161$ |
| | $N_v$ | | Total number of video lectures in the course. $N_v = 78$ |
| | $N_{a,max}$ | ► only | Maximum number of times a student in our course has clicked on the link for the same video to stream it online repeatedly. $N_{a,max} = 25$ |
| Basic Quantities | $i$ | | Student ID. It takes values from 1 to $N_s$. |
| | $j$ | | Video number in the assigned order. It takes values from 1 to $N_s$. |
| | $k$ | ► only | Index of streaming sessions. It takes values from 1 to $N_{a,max}$. |
| | $t$ | ► only | In-video time in seconds |
| | $A_{ijkt}$ | ► only | How many times student $i$ has **A**ccessed the content at in-video time $t$ of video $j$ during her $k$th streaming session of the given video |
| | $D_{ij}$ | ↓ only | How many times student $i$ has **D**ownloaded video $j$. |
| Derived Quantities | $TS_{ij}$ | ►&↓ combined | **T**otal number of **S**essions: how many times student $i$ has clicked on the link for video $j$ to either stream it online or to download it. |
| | $US_{ij}$ | ►&↓ combined | **U**nique **S**ession: whether student $i$ has ever clicked on the link for video $j$ to stream it online or to download it. $US_{ij}$ is either 0 or 1. |
| | $V_i$ | ►&↓ combined | Fraction of unique **V**ideos accessed: out of all 78 videos in the course, how many different videos has student $i$ ever accessed. Take that number and divide it by 78. |
| | $C(V)$ | ►&↓ combined | Complimentary cumulative distribution function of the amount of unique videos accessed by students |
| | $NS_j$ | ►&↓ combined | **N**umber of **S**tudents who have ever accessed a given video $j$ through either online streaming or downloading |
| | $FS_j$ | ►&↓ combined | **F**raction of **S**tudents that have ever accessed a given video $j$ through either online streaming or downloading (out of all students in the course) |
| | $FMS_j$ | ►&↓ combined | **F**raction of students with **M**ultiple accessing **S**essions for a given video $j$ (out of all the students who has accessed the given video at least once) |
| | $TA_{ijt}$ | ► only | **T**otal number of **A**ccesses for specific in-video content by a single student: for a selected video $j$, how many times has student $i$ accessed the content at $t$ second of the given video (summed over all video-streaming sessions made of the given video by the given student) |
| | $UA_{ijt}$ | ► only | **U**nique **A**ccess for specific in-video content by a single student: for a selected video $j$, whether student $i$ has ever accessed the content at $t$ second of the given video. $UA_{ijt}$ is either 0 or 1. |
| | $FC_{ij}$ | ► only | total **F**raction of **C**ontent in video $j$ that student $i$ has ever accessed |
| | $FSE_j$ | ► only | **F**raction of **S**tudents accessing almost the **E**ntirety of video $j$ (out of all the students who has ever accessed the given video through online streaming) |

Given that students did not access all videos, it is natural to ask whether a video's placement in the course had an effect on its corresponding $FS_j$ (i.e., fraction of students who accessed the given video j). For each video, the time when it was assigned, and the corresponding $FS_j$ are shown in Figure 5. Videos assigned in the first week had high fraction of students accessing, with the majority of them being accessed by more than 80% of the students. As the semester progressed, many students stopped accessing the lecture-oriented videos. For example, the fraction of students accessing lecture-oriented videos dropped to lower than 40% for the last batch of lecture-oriented videos assigned in the course. However, the fraction of students accessing laboratory videos did not seem to be affected by the videos' placement in the course. Even for the last 2 laboratory videos, the fraction of students accessing them remained higher than 84%.

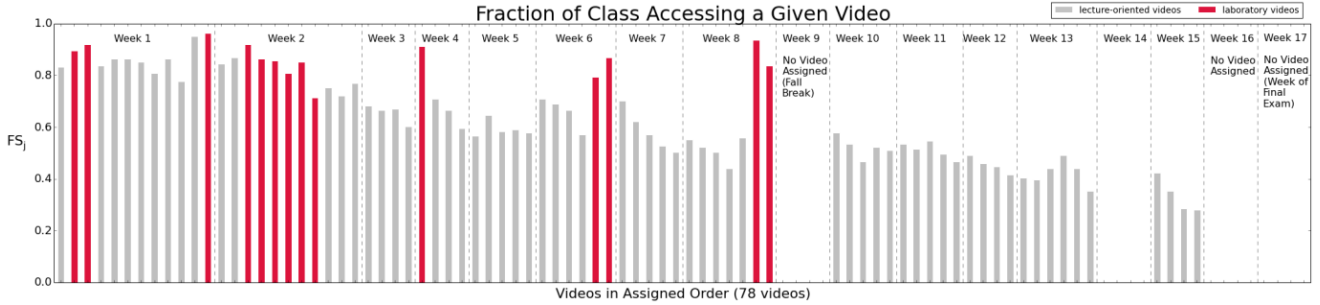

Figure 5. Fraction of accessing students ($FS_j$) for each video. The lecture-oriented videos are represented in gray, the lab-specific videos are represented in red, and the lab-supplemental videos are represented in pink. No videos were assigned in week 9, week 14, week 16 and week 17. The figure shows that students reduced their accessing of lecture oriented videos as time progressed while maintain their access for laboratory videos.

Table III. Descriptive information about the fraction of students with multiple accessing sessions ($FMS_j$) for lecture-oriented videos and laboratory videos.

|  | Lecture-Oriented Videos | Laboratory Videos |
|---|---|---|
| Maximum $FMS_j$ | 0.86 | 0.95 |
| Minimum $FMS_j$ | 0.19 | 0.46 (0.73 for lab-specific videos) |
| Median $FMS_j$ | 0.41 | 0.75 (0.90 for lab-specific videos) |
| Percentage of videos with $FMS_j \geq 0.5$ | 11% | 86% (100% for lab-specific videos) |

In addition to the fraction of accessing students, another difference found between the "lecture-oriented videos and "lab videos" lies in how often a single video in each group was accessed repeatedly by the same student. Table III shows the descriptive information about the fraction of students having multiple accessing sessions ($FMS_j$) of the same video for videos in the lecture-oriented group and the laboratory group. While for lecture-oriented videos, usually less than half of the accessing students would access these videos repeatedly, most students accessed the same lab video (especially the lab-specific ones) multiple times. Moreover, the behavior of accessing laboratory videos repeatedly continued throughout the semester. Overall, the results suggest that students engaged in laboratory videos (especially the lab-specific videos) much more than the lecture-oriented videos. We note that in the blended course, although students were instructed to watch the assigned videos before coming to the in-class section, they were not graded on whether or not they accessed the videos. Without direct grade incentives, students' video accessing behaviors were most likely driven by their personal interests. In this course, no instruction on how to perform the laboratories was provided in class; therefore, students may have high interests in the laboratory videos because these videos are considered the major resource helpful for completing the laboratory assignments. On the other hand, the lecture-oriented videos may not be viewed as the only resource helpful for learning about mechanics. For example, students may feel that their prior exposure to physics content was sufficient to learn the materials, and (or) that other course components (such as the in-class problem solving sections as some students mentioned in the course feedback) were more beneficial for learning the materials. In addition, the fact that the quizzes in the course were conducted on computers in which 100 submissions were allowed may have also contributed to the low number of accessing sessions for lecture-oriented videos. As a student responded in the end-of-course survey, "I will be honest. I only watched the lectures during the beginning of the course. They were helpful, but there was little motivation to watch them especially with how easy the weekly quizzes were." If no direct grade incentives were provided, and if students do not see a direct relation between how accessing lecture-oriented videos can help improve their performance in class, they may be likely to gradually disengage in these videos.

### 2. Timing of Video Access

In addition to the number of accessing sessions students made of each video, of equal interest is the time when students accessed videos. Figure 6 plots the total number of video-accessing sessions made by students (i.e., $\sum_{ij} TS_{ij}$) on each day in the semester. Figure 6 shows that the temporal profile of lab video accessing did not correlate with the profile for lecture-oriented videos, suggesting that students tend to access different types of videos at different times. Overall, the lab videos were accessed most frequently the week before the laboratory due dates (although these videos had been assigned at least one week earlier, as described in the caption of Figure 1). In particular, a good number of lab-video accessing sessions were made on the day when the labs were due [19]. The lecture videos, on the other hand, were accessed each week. In addition, lecture videos were accessed frequently on weekdays and infrequently on weekends, except for the Saturday and Sunday right before the midterm exam. In this course, in-class problem solving sections were held on Mondays to Thursdays, weekly quizzes were held on Fridays, and weekly homework assignments were due on Sundays [20]. This structure may have contributed to the higher number of accessing sessions for lecture-oriented videos in the weekdays (for example, our anecdotal experiences suggest that some students would access videos during the in-class problem solving sessions). In addition, there was a spike in lab video accessing sessions on lab due dates, but there was no spike in accessing sessions for lecture-oriented videos on homework due dates. This may imply that the lecture-oriented videos were not the primary resource students resorted to when completing the homework assignments. Other components in the course, such as the in-class problem solving sections as mentioned in student responses to the end-of-course survey, may be considered more relevant for homework completion by the students.

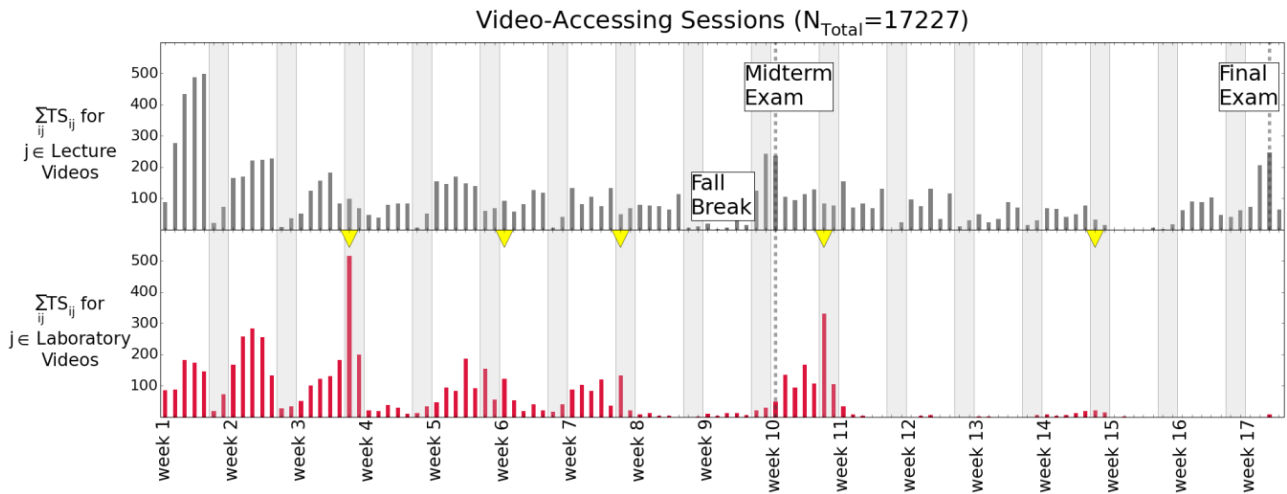

Figure 6. Total video accessing sessions (separated into laboratory videos and lecture-oriented videos) made by all students on each day in the semester. If a student accessed the same lecture-oriented video twice on the same day, or if she accessed two different lecture-oriented videos each by once, in both cases, her contribution for the lecture-oriented video-accessing sessions made on that particular day is 2. In this figure, weeks start on Mondays, and shaded regions indicate weekends. Laboratory due dates are demarcated by yellow triangles along the top horizontal axis. The last batch of videos were assigned in week 15. Quizzes were given every Friday from week 1 to week 14 (except for week 9 due to fall break). There was generally one homework due per week, with the last homework assignment due in week 16. The final exam was scheduled on the Wednesday in week 17. For the few students who had a conflicting final exam schedule, their final exam was held on the Thursday in week 17. Overall, students accessed videos more prior to lab due dates, quizzes and exams.

While Figure 6 provides a global view of when students accessed videos, we can look more in-depth into which exact videos students accessed during a particular period of time to obtain a more thorough understanding of students' video accessing behaviors. Figure 7 plots the total number of unique students who accessed a given video (i.e., $NS_j$ for given video $j$) on each day in the semester. Figure 7 shows that when accessing videos, students typically focused on videos assigned in the current week or the week before. Although there were a few revisits of prior videos when the midterm and final exam came, the number of students accessing old videos was not high. In total, only about one-fifth of the students (21.7% for midterm exam and 23.6% for final exam) revisited at least one old video during the exam period. (Here, old videos are defined as videos that had been accessed earlier in the semester and were assigned more than 2 weeks ago.) This finding is consistent with our previous finding that the majority of videos were only accessed by students only once. It also echoes a prior study in a Circuits and Electronics MOOC [6], which shows that lecture videos are resources that MOOC students spent the most time on when preparing for weekly homework, but not for preparation of exams. Students in our blended course were in some sense similar to the MOOC students in the prior study in that compared to exams, their video usage seem to relate more to recent weekly work (such as in-class problem solving sessions, homework assignments or quizzes that typically cover content from videos assigned in the current week or the week before).

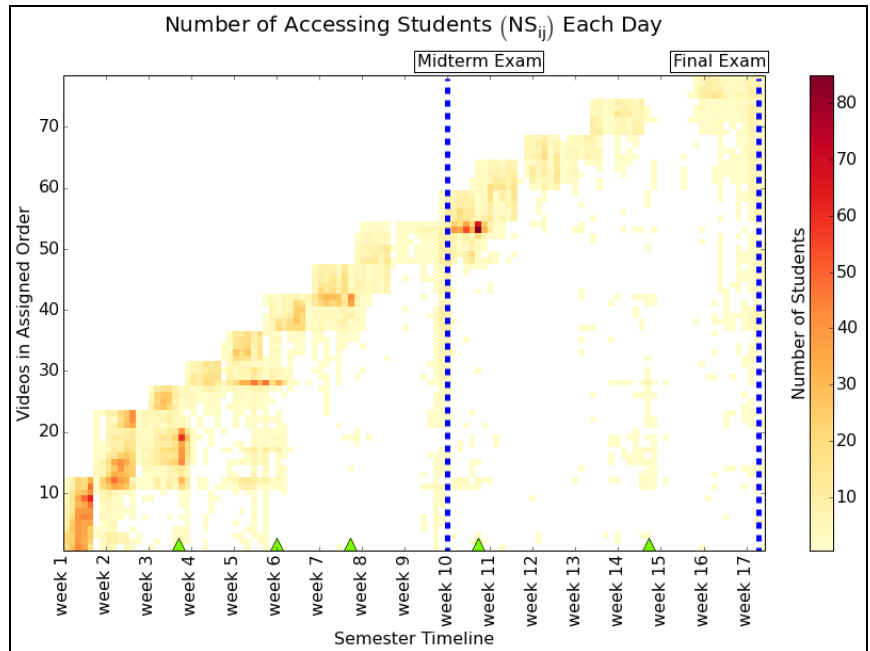

Figure 7. Total number of students accessing a given video ($NS_j$) on each day in the semester. Videos in this course were assigned by weeks. As the semester progressed, new batches of videos were made available when new weeks came, resulting in stair-step behavior observed along the diagonal line of the figure. For example, videos for the 2nd week (i.e., videos number 13 to 23) were not available to the students until a few days before week 2; therefore, there was no student accessing these videos in the first five days of week 1. Once videos were assigned, they remained available throughout the rest of the semester. Lab due dates were demarcated with the green triangle on the horizontal axis. Overall, the figure shows that students rarely revisited videos that were assigned from previous weeks (except for few revisits immediately before major assessments such as the midterm and final exams).

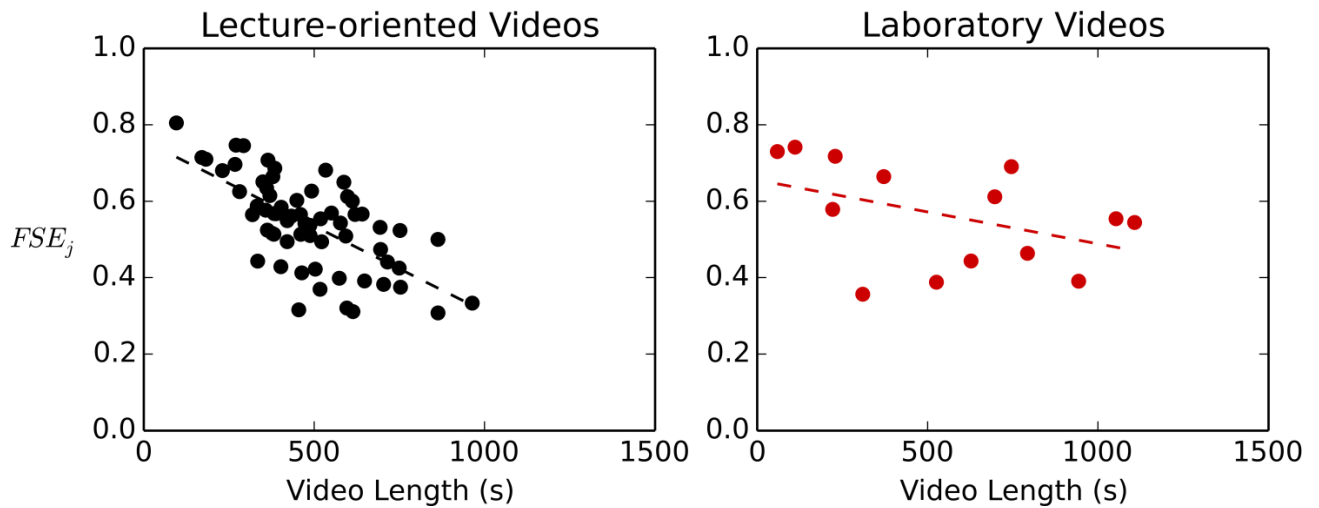

Figure 8. Video length vs. fraction of students accessing more than 99% of the given video ($FSE_j$). The trendlines for lecture-oriented videos and laboratory videos are y=-0.000447x+0.758 and y=-0.000167x+0.655, respectively. The figure shows that compared to laboratory videos, students were more likely to not access the entirety of lecture-oriented videos when these videos became longer.

### 3. Fraction of In-video Content Accessed

With a basic understanding of how students accessed videos (e.g. which types of videos drew their attention more, and when students accessed videos), we now take a step further to investigate how students interacted with the videos after they clicked on the video link. In particular, we focused on the fraction of in-video content accessed by students and whether students would access the entire video once they decided to access the given video. Figure 8 shows the fraction of students who accessed at least 99% of a given video ($FSE_j$) and how this quantity relates to the length of the given video. When averaged over all videos in the group, the mean FSE is 0.54 for lecture oriented videos and 0.56 for laboratory videos, suggesting that when students accessed videos, only half of the time they would access almost the entirety of the video. In addition, Figure 8 shows that when lecture-oriented videos become longer in length, the fraction of students accessing almost the entire video decreases with a slope of -0.00045 per second. For the lab videos, however, the fraction of complete-accessing students did not decrease as much with video length (slope=-0.00016 per second). The finding suggests that for these laboratory videos, which typically contain specific information necessary for completing laboratory work, students not only accessed them more frequently but also were more likely to access the entirety of these videos despite their lengths. A higher interest in the laboratory videos than lecture-oriented video was therefore not only observed in the video-accessing session level but also at the detailed-interaction level.

### 4. Relation Between Video Accessing and Student Learning

Table IV. Relation between video accessing and student performance. The Pearson Correlation Coefficient (r) for each case is provided. Overall, students' course performance does not correlate to their video accesses.

| Video accessing vs. Student performance | r |
|---|---|
| Number of videos accessed vs. final scores | 0.16 |
| Total time spent accessing videos vs. final scores | 0.00 |
| Frequency of pause interactions (i.e., total number of pauses divided by total unique in-video times that have been accessed) vs. score on final exam | 0.12 |
| Time spent accessing lab videos vs. grades received on related lab | 0.11 |
| Average fraction of video a student accesses vs. final score | 0.28 |
| Total unique in-video time accessed vs final score | 0.23 |
| Number of videos accessed vs. incoming GPA | 0.10 |
| Total time spent accessing videos vs. incoming GPA | 0.12 |

With a basic understanding of students' video accessing behavior, we now shift focus to the relation between video accessing and student learning. Table IV shows the correlation between video accessing and student performance in the course. Although a number of different measures of student video accessing behaviors are used (e.g., the number of videos accessed, total time spent accessing videos, total length of videos accessed by students, frequency of interactions with video player), overall, none of them shows a strong correlation to student performance in the course.

### 5. Detailed student interaction with videos

As presented in the previous section, in our course, students' performance did not correlate to their video accesses. A prior study [12] suggests that if students had a surface approach to learning that focused on the concrete aspects of a task rather than the meaning of the task, then more engagement with video lectures would not necessarily lead to better performance. In order to get a deeper insight into how students in our course interacted with the videos in detail and what type of video content (if any) students in our course seemed to engage more with, an exploration involving the most-accessed laboratory video ($j$=15) and the most-accessed lecture-oriented video ($j$=11) was conducted. The findings suggest that students in our blended course seemed to pay more attention to content that provides *concrete* information useful for assignment completions. However, students may not engage as much with other content that is also considered important from an instructor's point of view. For example, Figure 9 shows the total number of accesses made by students for each particular second in the most-accessed laboratory video (i.e., $\sum_{i=1}^{N_S} TA_{ijt}$, $j$=15). In addition to the number of accesses, we took pausing as another indicator of student interest, since we would expect a student to pause a video to take notes or to repeat important passages in the video. Places where students paused the most are plotted in Figure 9, too. Overall, the high-frequency noise shown in the number of accesses in Figure 9 suggests that students skipped frequently through the video. Moreover, Figure 9 indicates that in this laboratory video, which teaches students how to construct a computational model for constant velocity motion from a starter python code provided, high student access occurs at places where *actions* useful for completing the corresponding laboratory assignment are demonstrated on the screen. (See Table V for a description of the *action* each arrow represented in Figure 9.) Here, we define *action* as a situation in which a physical act (such as downloading a starter python code, entering a parameter value required, running the code to check for the result) from the student is required or explicitly recommended in the video for the completion of the assignment. If, for example, the video discusses an important parameter in the python code, but there is no need to change the default value in the starter code for that parameter, it does not constitute an action defined here. Figure 9 shows that students not only accessed these "*action*" sections more frequently but also paused a lot in these sections. On the other hand, sections in which the instructor discusses the physics concepts behind but no modification of the code is required are less engaged by the students. For example, from 600s to 673s, the instructor points out that in the iteration loop where each time the motion would be predicted a small time Δt into the future using Newton's 2$^{nd}$ law ($\frac{\vec{F}_{net}}{m} = \frac{\Delta \vec{v}}{\Delta t}$), Δt needs to be chosen such that it is much less than the typical time scale of the motion being observed. The instructor discusses that for the motion of the ball being observed, which lasts about 1 second, the default value Δt provided in the starter code, which is 1/100 times smaller than 1 s, is suitable. He further points out that it is not necessary to set Δt to be equal to the time between frames of the recorded motion, especially when the time between frames is large. While this discussion contains important physics behind the computational model, however, students did not seem to engage with this section as much as they do with the "action" sections.

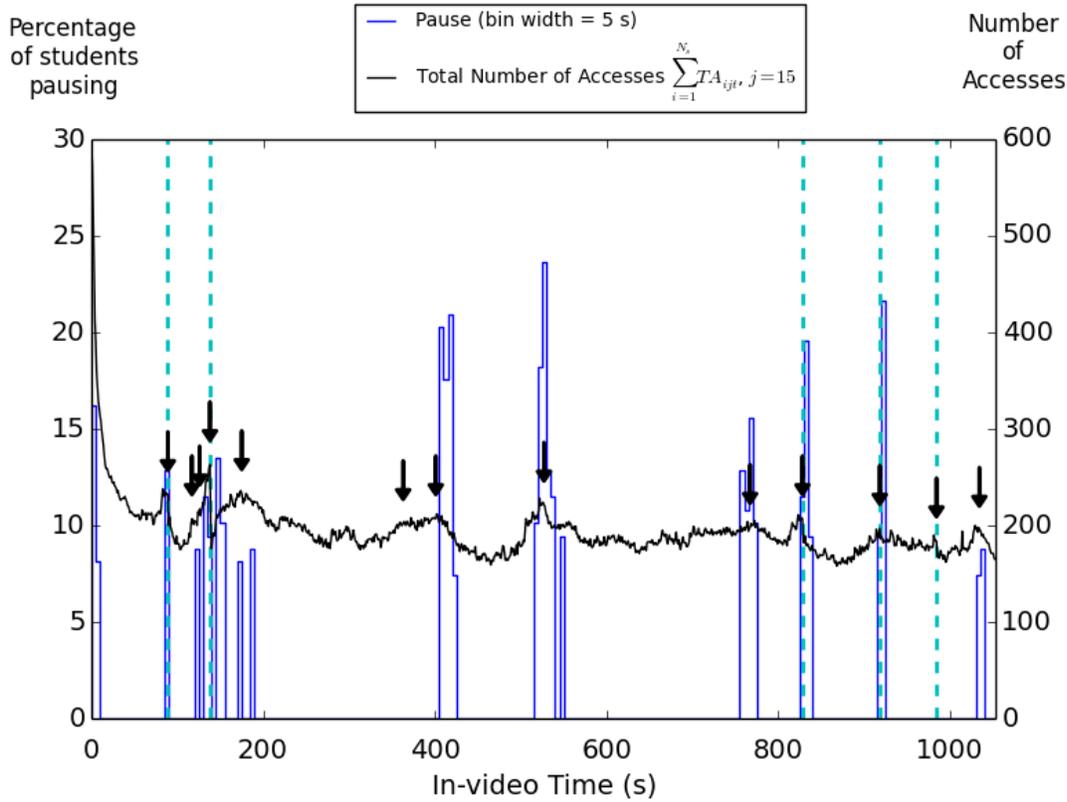

Figure 9. Clickstream analysis of video 15. This video, entitled "Creating a Computer Model of Constant Velocity Motion", was accessed by 148 unique students in total. The black line represents the number of times each particular point of a given video was accessed by students (i.e., $\sum_{i=1}^{N_s} TA_{ijt}$ where $j=15$). The blue line represents the percentage of unique students who has ever paused within each 5-second time window. Since we are interested in student-generated pauses, the computer-generated pauses both at the interaction points and the very end of the video have been taken out. In order to help identify peaks, only pauses which is equal to or higher than two median absolute deviations [21] ($2\sigma$) above the median, which in this case corresponds to 7.4%, are shown in the figure. The interaction points in the video are indicated by the vertical dashed line. Places where "actions" are demonstrated in the video are indicated by the black arrows. The figure shows that student accesses increased during these action points. A description of the action each arrow represented is included in Table V.

Table V. Description of the action each arrow represented in Figure 9

| In-video time | Actions demonstrated/suggested in the video |
|---|---|
| 89s | Watch a pep talk on coding and computing using the link provided |
| 117 s | Launch VIDLE (a software application that provides the setting where codes will be written and run) |
| 126 s | Open the starter code |
| 138 s | download the starter code using the link provided if students student had not done so |
| 175 s | Run the unedited starter code to make sure it runs without error |
| 363 s | Rotate the orientation of the visualization window by right clicking and dragging |
| 401 s | Edit a line of code so that it correctly represents the mass of the object |
| 527 s | Edit the codes so that the initial conditions of the object are correctly included |
| 767 s | Edit a line of code to specify the iteration limit of the while loop |
| 828.5 s | Edit a line of code so that it correctly represents how velocity update should be performed |
| 918 s | Edit a line of code so that it correctly represents how position update should be performed |
| 984 s | Edit a line of code to put in the net force for the motion under study |
| 1034 s | Run the program |

Figure 10 shows the total number of accesses and the percentage of students pausing at various in-video time for the most-accessed lecture oriented video (*j*=11) For this video entitled "Newton's 2$^{nd}$ Law", a slight increase in the number of accesses was found at the points where examples of net force vector sums (185 s) and Newton's 2$^{nd}$ Law are displayed (277 s). Pauses are found in these regions too, as well as a couple of other regions that discuss *concrete* information or techniques useful for problem solving (such as 75~80 s where the instructor discusses the dimension and SI unit of mass; 110~115 s where the instructor discusses the fact that forces have magnitude (i.e., how strong?) and direction (i.e., which way?); 235~245 s where the instructor discusses a tip for finding net force; see Table VI for a full description of the video content that corresponds to each major pause peak in Figure 10). While these sections contain important information, other sections also convey important concepts associated with Newton's 2$^{nd}$ Law. For example, from 308~324 seconds, the instructor discusses the important role of Newton's 2$^{nd}$ Law $\frac{\Delta \vec{v}}{\Delta t} = \frac{\vec{F}_{net}}{m}$ by pointing out

"*Newton's 2$^{nd}$ law is an amazing statement which relates something we can obtain just by directly watching our object moves during some time interval, to, very different quantities that we cannot generally get at just by looking at our object*".

From 363~417 s, the instructor discusses the epistemology behind Newton's second law and explains what makes Newton's 2$^{nd}$ law a law by saying

"*First, Newton's second law is a law because it tells us a secret of the universe. It's not obvious that quantities that describe motion should be related to object properties with the influence of the surroundings this particular way. Second, Newton's 2$^{nd}$ law is a law rather than somebody's opinion or a wild guess, because it is a statement about nature that has withstood the tests of countless experiments with moving objects over a wide variety of conditions. This means you don't have to believe it just because I told you so. You'll have plenty of opportunities to check this for yourself*".

However, students accessed and paused less at these sections, suggesting that students may not focus as much in such type of information that is typically less explicitly manifested in a problem solving process despite the importance of this information in the construction of a solid understanding of physics. In fact, when students were asked about their feedback for the lecture videos in an end-of-course survey, a common opinion expressed by the students is that the videos would be more helpful if they focus on the applications more. For example, a student points out that "*It would be helpful if the videos made it clearer which formulas were important. For example, at the end of each video all the different formulas introduced in the lecture could be written on the screen.*" Another student points out that "*a lot of the time they [the videos] explained the concepts really well. Which is great and all but they didn't always address how to apply the concepts, which is all that really matters in this course. If you understand the concept, but you can't do the problem, you still get the problem wrong.*" Other responses mentioning "*more examples*" are also frequently found.

In Figure 10 and

Table VI, it is also worth noting that the peak corresponding to mathematical manipulation (in this case dimensional analysis) has the highest number of students pausing among all the pauses. While other information such as tips for finding net force (e.g. never count forces on the surroundings; do not include forces that the objects exert on themselves) may be considered as important as dimensional analysis from an instructor's point of view, there are more pauses for the latter (285~300 s) than the former (230~245 s). Although these tips from 230~245 s were designed based on common student difficulties related to finding the net force, students may not necessarily be aware of the importance of this discussion. Therefore, they may not feel the need to pause there as much as they do for places in which mathematical equations appear on the screen.

The findings above echo a prior study which shows that experts and students may view the same physics videos differently [22]: while instructors may believe that the answer to a physics question has been addressed in a given video, students may not necessarily think so. Similarly, based on our exploration of student-video interaction with two videos in the course, it is likely that while instructor believes a concept has been emphasized in the videos, students may not necessarily notice the importance of it and/or pay attention to it. Moreover, a glance through students' in-video content accessing behavior of all 78 videos indicates an interesting finding: the total number of accesses (i.e., $\sum_{i=1}^{N_s} TA_{ijt}$) usually increased toward an interaction point and dropped immediately after the interaction point (see the 1st, 2nd, 3rd, and 5th interaction points in Figure 9 for examples). In the 86 interaction points given in the videos throughout the course, 66 of them had a higher $\sum_{i=1}^{N_s} TA_{ijt}$ compared to that of 5 seconds before, and 60 of them had $\sum_{i=1}^{N_s} TA_{ijt}$ that was lowered by at least 10% at 5 seconds later. Since a question for students to answer is usually posed at an interaction point, and the instructor would typically discuss the given question in details right after the interaction point, this finding suggests that students seemed to be interested in accessing these questions. However, although these questions usually draw students' attention, some students may be satisfied once they obtained the correct answer to the question and they were less interested in exploring the concepts behind the question further. All these findings suggest that there may be a potential for students to benefit from the videos more if the videos are placed in the context of a larger course structure with additional instructional strategies, guidance, and/or activities provided to help students contemplate and organize the information presented in the video in more depth.

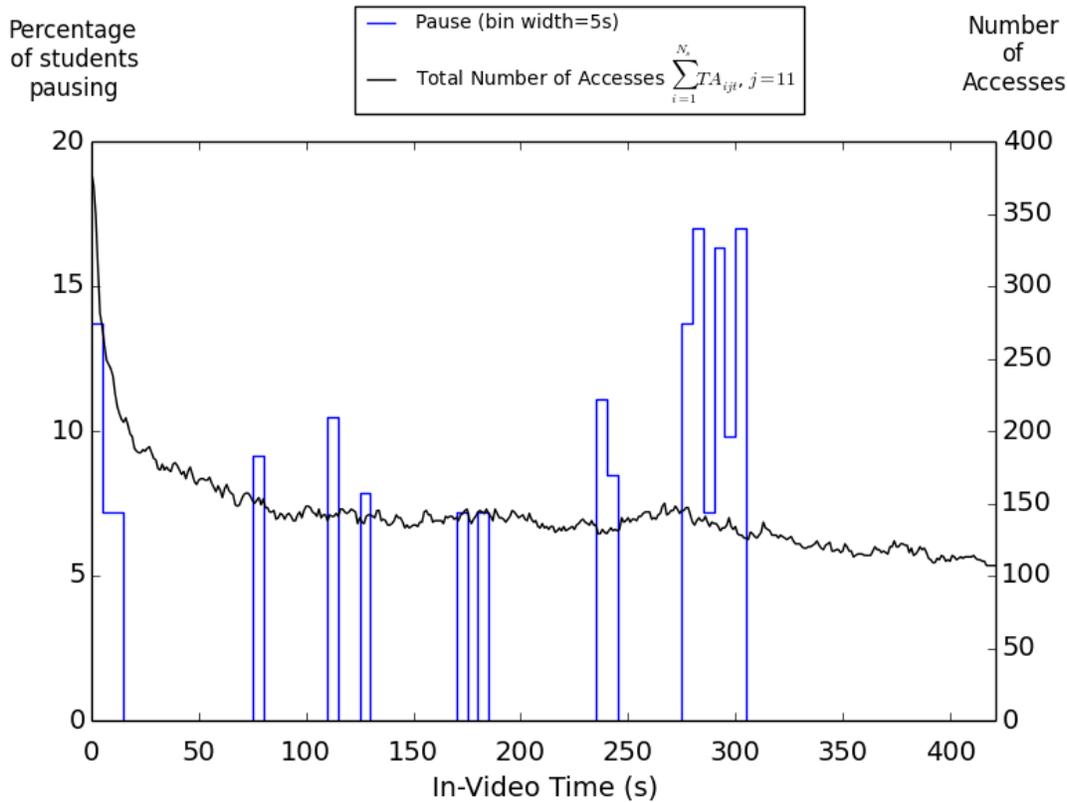

Figure 10. Clickstream analysis of video 11. This video, entitled "Newton's 2$^{nd}$ Law", was accessed by 153 unique students. The black line represents the number of times each particular point of a given video was accessed by students (i.e., the value at in-video time t was obtained by $\sum_{i=1}^{N_s} TA_{ijt}$ where $j$=11). The blue line represents the percentage of unique students who has ever paused within each 5-second time window. To help identify common pause peaks, only pauses which is equal to or higher than two median absolute deviations (2σ) above the median (which in this case corresponds to 7.2%) were shown in the figure. Since we are interested in student-generated pauses, automatic pauses occurred at the very end of the video have been taken out.

Table VI. Description of the video content that corresponds to the major pause peaks in Figure 10.

| In-video time | Description of video content |
|---|---|
| 75~80 s | Dimension and SI unit of mass |
| 110~115 s | Forces have magnitude (how strong?) and direction (which way?) |
| 125~130s | (When drawing an arrow to represent a force) the length of the arrow represents not a length but the strength of the force. |
| 170~185 s | Finishing the statement that net force is the sum of all forces acting on the system and demonstrating how to perform vector sums to find the net force when each push or pull is represented by a vector. |
| 235~245 s | 3$^{rd}$ tips for finding net force: objects do not exert forces on themselves (while info for the 1$^{st}$ and 2$^{nd}$ tips is still on the screen; tip 1: count all forces from surroundings on system tip 2: never count forces on the surroundings) |
| 275~305 s | • Mathematical Expression for Newton's 2$^{nd}$ law; <br> • Discussion that both Fnet and delta v are vectors, i.e., they have direction and magnitude <br> • units on both sides of the Newton's 2$^{nd}$ law equation <br> • using dimensional analysis to find the unit of net force; a table summarizing the dimension and SI unit of force |

## VI. DISCUSSION AND CONCLUSIONS

In this study, student engagement of video lectures in a blended introductory mechanics course in which in-class lectures were replaced by online lecture videos was explored. In this course, students were not graded on their video accesses, and our findings suggest that students may use the videos based on perceived value. For example, in this course, little in class time was spent describing or working on laboratory activities, and all instructions required for laboratory activities were delivered through lab videos (especially the lab-specific ones). Students not only accessed the laboratory videos a lot, but also were more likely to access these videos completely without skipping sections in it when the video length increases. On the other hand, the fraction of students accessing lecture-oriented videos decreased to less than 40% toward the end of the semester, and students were more likely to not access the whole video when these lecture-oriented video becomes longer. The low access for lecture-oriented videos (as measured by $C(V)_{lec}$, $FS_j$ and $FSE_j$) may be due to factors such as students' prior exposure to physics content precluded the need to access the lecture-oriented videos, and/or that students feel that solving problems in class was sufficient (or more helpful) for learning the materials. The end-of-course survey also indicates that students may tend to pay more attention to how to apply the physics concepts/principles learned, but the lecture videos did not provide as many examples as they would hope - which could be another reason why students are less motivated to access videos. The fact that most students could easily perform well on the weekly quizzes, which were conducted on computers in which 100 submissions were allowed, may have also contributed to the low $C(V)_{lec}$, low $FS_j$ and low $FSE_j$ for lecture-oriented videos.

Our study also found little correlation between student engagement with video lectures and student performance in the course (see Table IV). This result is similar to our experiences from several semesters of traditional lecture-style introductory mechanics courses held at the Georgia Institute of Technology, in which weak correlation (r=0.33) between student attendance of lecture and student performance on final exam was found. However, with the advantage of having the lectures held online in the blended course, we were able to explore in-depth how students interacted with different content of the lectures in an unprecedented way. An examination of the content that students focused on from two selected videos – the most accessed laboratory video and the most accessed lecture video – suggested that students may engage more with concrete information that is explicitly required for assignment completion (e.g., actions required to complete laboratory work, or formulas/mathematical expression that are typically manifested explicitly in a problem solving task). Other types of content, such as the underlying physical implication of a principle or the epistemology behind, seemed to be less engaged.

Since our current study that looked more in-depth into students' in-video interactions focused primarily on two example videos, future work that extends this line of research to all videos in the course can be conducted to obtain a more holistic view of how students engage with videos. These results can have great potential to help shape the instructional designs of the videos and/or the course structure to better support student learning. For example, since our findings suggested that students' in-video accesses usually increased toward interaction points, instructors could consider adding more interaction points in regions where important concepts were found to be frequently ignored by students. In-class time could also be spent more effectively to target common student difficulties. We note that while places students engaged the most in a video may indicate what students consider to be important content in the course, when the materials covered in these regions are complicated, the increased number of student accesses or pausing events may also indicate points of confusion. Using clickstream data (possibly with the help of a short quiz or survey) can help identify these points of confusion so that in-class time could be allocated more effectively. Moreover, instructors could also consider adopting a different instructional approach/example than that employed in these points of confusion to better assist students with their learning. These results can also inform the instructors how to revise their videos to better fit their instructional goals. If students were not necessarily motivated to access videos, external intervention (such as instructors explaining why videos are important components in the course, discussing how students could best use these videos to benefit their learning, and providing activities to help students see the benefit of accessing videos) may be implemented so that this resource can be put in more effective use. In sum, the clickstream data provides us with a powerful tool to obtain a deeper insight into how students made use of online lectures in the course. With iterative modification of the videos and/or course designs based on implications from these data, instructors can construct a more effective learning environment to better suit their instructional goals.

## VII. ACKNOWLEDGEMENTS

This project was funded by the Bill and Melinda Gates Foundation, the Georgia Governor's Office of Student Achievement, and the National Science Foundation (NSF DUE-0942076).

## APPENDIX:

### Constructing matrix $A_{ijkt}$ from clickstream data

Three types of events- play, pause, and seek- are involved in the construction of matrix $A_{ijkt}$. Since $A_{ijkt}$ represents how many times student $i$ has accessed the content at in-video time $t$ of video $j$ during the $k$th time she streamed this video, every time student $i$ clicked on play at in-video time $t_1$ (in second) of video $j$ and then clicked on pause after the video proceeds to in-video time $t_2$ (in second), a value of 1 is added to all elements between $A_{ijkt_1}$ and $A_{ijkt_2}$. Streaming sessions in which seek events are involved require more work. Students can, for example, watch the first 40 seconds of a video, and then skip to the point that corresponds to 100 s in in-video time, and then continue watching the video for another 20 second. Depending on whether the student had clicked on "pause" before she skipped to t=100 s in the video, two patterns of recorded events are possible. The first type of pattern shown in Table VII corresponds to the case in which students had clicked on "pause" before she skipped away from t=40 s. The second type of pattern shown in Table VIII corresponds to the case where students did not clicked on "pause". We note that at the time when our course was offered, Coursera only recorded when a "seek" event ended, but not when it started. In the second case where a student skipped to a different point in the video while the video was still playing (i.e., the pause button had not been clicked), an estimation of the portion of video accessed before the seek event happened is made by multiplying the difference in real time between the seek event and the previous event by the average playback rate [23] student used. In the example shown in Table VIII, it is estimated that the portion of video accessed between the first play event and the 2$^{nd}$ seek event is t=0s to t=0+40*1=40s.

Table VII. Example made-up data explaining how events would be recorded in clickstream (the first 3 columns [24]) and how matrix $A_{ijkt}$ would be constructed (the last 3 columns) for the following scenario: a student watched the first 40 seconds of a video, clicked on pause, and then clicked on the scrubber below the video to jump to the point that corresponds to 100 s in in-video time. Then, she clicked on play to continue watching the video for another 20 second. In this example, the playback rate student used was 1.

| Event | Real time | In-video time | Implication for elements in matrix $A_{ijk}$ | | |
|---|---|---|---|---|---|
| Play | 0 s | 0 s | A 1 is added to every element between $A_{ijk0}$ ~ $A_{ijk40}$ | | |
| Pause | 40 s | 40 s | | | |
| Seek | 41 s | 100 s | | None | |
| Play | 42s | 100 s | | | A 1 is added to every element between $A_{ijk100}$ ~ $A_{ijk200}$ |
| Pause | 62 s | 120 s | | | |

Table VIII. Example made-up data explaining how events would be recorded in clickstream (the first 3 columns) and how matrix $A_{ijkt}$ would be constructed (the last 3 columns) for the following scenario: A students watched the first 40 seconds of a video, and then (without clicking on "pause" first) clicked on the scrubber below the video to jump to the point that corresponds to 100 s in in-video time. Since the pause button had not been clicked, the video would automatically continue playing from t=100 s. The student then continued watching the video for another 20 second. In our study, we assume that the real-time it takes for the video to jump from in-video time t=40 s to t=100 s is negligible. The playback rate student used was 1.

| Event | Real time | In-video time | Delta Real time | Implication for the elements in array $A_{ijk}$ | |
|---|---|---|---|---|---|
| Play | 0 s | 0 s | -- | $A_{ijkt}$ =1 for $t$ in [0, 0+40*1] | |
| Seek | 40 s | 100 s | 40 | | $A_{ijkt}$ =1 for $t$ in [100, 120] |
| Pause | 60 s | 120 s | 20 | | |